\documentclass[a4paper,fleqn,usenatbib,useAMS]{pazh_mnras}

\usepackage{graphicx}	% Including figure files
\usepackage{amsmath}	% Advanced maths commands
\usepackage{amssymb}	% Extra maths symbols
\usepackage{multicol}        % Multi-column entries in tables
\usepackage{bm}		% Bold maths symbols, including upright Greek

\usepackage{color}
\usepackage[T1]{fontenc}
\usepackage{ae,aecompl}
\usepackage{newtxtext,newtxmath}

\newcommand{\msun}{M_{\sun}}

\newcommand{\Mtot}{M_{\rm tot}}
\newcommand{\Mism}{M_{\rm ISM}}
\newcommand{\re}{r_{\rm eff}}
\newcommand{\ergs}{\,erg\,s$^{-1}$}

\def\2MASS{\textit{2MASS}}

\def\XMM{\textit{XMM-Newton}}
\def\Chandra{\textit{Chandra}}
\def\ROSAT{\textit{ROSAT}}
\def\GALEX{\textit{GALEX}}

\newcommand*{\PathtoFigs}{IMAGES}
\newcommand\Ngal{13}

\title[Elements Diffusion in the ISM]{Diffusion of Elements in the Interstellar Medium in Early-Type~Galaxies}

\author[Medvedev, Sazonov \& Gilfanov]{\bf P.~Medvedev$^1$\thanks{E-mail: \href{mailto:tomedvedev@iki.rssi.ru}{tomedvedev@iki.rssi.ru}}, S.~Sazonov$^1$ and M.~Gilfanov$^{2,1}$\\
$^{1}$Space Research Institute (IKI), Profsouznaya 84/32, Moscow 117997, Russia\\
$^{2}$Max Planck Institute for Astrophysics, Karl-Schwarzschild-Strasse 1, 
85741 Garching, Germany\\
}

\pubyear{2017}

\begin{document}
\label{firstpage}
\pagerange{\pageref{firstpage}--\pageref{lastpage}}
\maketitle

\begin{abstract}
We consider the role of diffusion in the redistribution of elements in the hot interstellar medium (ISM) of early-type
galaxies. It is well known that gravitational sedimentation can affect significantly the
abundances of helium and heavy elements in the intracluster gas of massive galaxy clusters. 
%The self-similar behaviour of the temperature profiles of the relaxed cool-core clusters 
%the element sedimentation should be particularly effective 
The self-similarity of the temperature profiles and 
tight mass--temperature relation of relaxed cool-core clusters 
suggest that the maximum effect of sedimentation take place 
in the most massive virialized objects in the Universe. 
However, \Chandra\ and \XMM\ observations demonstrate
more complex scaling relations between the masses of early-type galaxies and other parameters, such as
the ISM temperature and gas mass fraction. An important fact is that early-type galaxies 
 can show both decreasing and increasing radial temperature profiles. We have calculated the diffusion based on the observed
gas density and temperature distributions for \Ngal\  early-type galaxies that belonging to the different environments and
cover a wide range of X-ray luminosities. To estimate the maximum effect of sedimentation and thermal
diffusion, we have solved the full set of Burgers' equations for a non-magnetized ISM plasma. The
results obtained demonstrate a considerable increase of the He/H ratio within one effective radius for all
galaxies of our sample. For galaxies with a flat or declining radial temperature profile the average increase of
the helium abundance is 60\% in one billion years of diffusion. The revealed effect can introduce a significant
bias in the metal abundance measurements based on X-ray spectroscopy and can affect the evolution of stars
that could be formed from a gas with a high helium abundance.
\end{abstract}

\section{Introduction}
\label{sec:intr}
The role of diffusion in shaping the spatial distribution
of elements in the intracluster medium (ICM) of
galaxy clusters has been the subject of many previous
studies \citep{Fabian1977,Gilfanov1984,Chuzhoy2004,Peng2009,Shtykovskiy2010,Medvedev2014}. 
In the approximation of a
non-magnetized ICM plasma the diffusion is driven
 by the gravity,
concentration and temperature gradients.
 In particular, large temperature
gradients in cool-core clusters slow down the
sedimentation of elements heavier than hydrogen at
the cluster center. Given the observed self-similar
ICM temperature profile in relaxed clusters \citep{Vikhlinin2006},
it can be asserted that the integrated effect of diffusion
depends mainly on the global properties of the
clusters: the total mass ($M$), the mean temperature
($T$), and the gas mass fraction ($f_b$). However,
these parameters are not independent. Observations
\citep{Vikhlinin2006} revealed a tight correlation
between the cluster temperature and total mass,
in agreement with expected on theoretical grounds 
\citep[see, e.g.,][]{Mathiesen2001}. At the same time,
the enclosed baryon fraction in clusters is expected to be close
to the cosmic mean
\citep{White1993}, which is also confirmed
by observations \citep{Vikhlinin2006}. The small
scatter of values in the relations suggests a similar
galaxy cluster formation history. Thus, one may talk
about an explicit dependence of the sedimentation
amplitude on cluster mass and expect a maximum
effect for the most massive (hot) virialized objects.

In this paper we consider the elements diffusion in the interstellar medium (ISM)
in early-type (elliptical and lenticular)
galaxies. Although the total mass, temperature,
and gas mass fraction are also correlated for such objects,
there is a large scatter at a fixed value of any
parameter. It is also important that no ``universal''
temperature profile is observed for early-type galaxies: the ISM can have both
decreasing and increasing with radius temperature profiles. 
This can affect significantly the sedimentation
amplitude.

The scatter in observed X-ray properties can be
partly related to the difficulty of studying the emission
from the ISM. 
%The X-ray emission of
%elliptical galaxies is often difficult to separate from
%the emission of the surrounding hotter intergalactic
%gas. 
It is often challenging to disentangle 
relatively weak ISM emission from the strong emission of surrounding ICM 
gas.
X-ray spectroscopy of low-luminosity galaxies
runs into the problem of a too small signal-to-noise
ratio. Yet another difficulty in studying such objects
is related to the necessity of separating the diffuse thermal
X-ray emission of the ISM from the
emission of the galactic stellar population (low-mass
X-ray binaries and cataclysmic variables) and occasionally
also the emission of the central supermassive black
hole. On the other hand, by now dozens of early-type
galaxies have been observed with \Chandra\ and \XMM, and, hence,
a large volume of sufficient quality data
has been accumulated. The unprecedented \Chandra\ 
angular resolution and the high \XMM\ sensitivity
allow the gas temperature and density to be
obtained in a uniform way for the entire sample of galaxies
being studied.

The mass--temperature relation demonstrates a
more complex behaviour for early-type galaxies \citep{Boroson2011} than does the virial relation,
$M \sim T^{3/2}$, which is closely followed by the measured relation
for galaxy clusters (adjusted for the same redshift). Besides,
the measured gas temperatures in such galaxies
always turn out to be higher than the virial values. 
This suggests the importance of additional gas heating
sources, apart from the heating by adiabatic compression
during infall in the galactic gravitational potential and the
thermalization of the gas kinetic energy associated
with random stellar motions \citep[see][]{Pellegrini2011}. The
explosions of SNe Ia and the accretion of gas onto
a central supermassive black hole can be such additional
heating mechanisms \citep[see, e.g.,][]{Ciotti1991}.
The contribution of additional heating sources to the
overall gas thermal balance changes with galaxy
mass. This is observed as a change of the slope in
the $\sigma_c-T$ relation for galaxies with different luminosities:
the faintest galaxies deviate more strongly
from the virial relation \citep{Pellegrini2011}. Note that
the central velocity dispersion, $\sigma_c$, is believed to be
a representative measure of the galactic potential well \citep{OSullivan2003}; therefore,
one may talk about similar features in the $M-T$
relation.

In the case of galaxy clusters, the baryon fraction
$f_b$ approaches the cosmological value within a radius
$r\sim r_{500}$(within which the mean density exceeds the
critical density of the Universe by a factor of 500).
Since the accessible region of observations of early-type
galaxies is considerably smaller than $r_{500}$, it is
hard to judge the universal value of $f_b$. Measurements
of the density of the ISM show a 
smaller gas mass fraction in the accessible
region of observations. This is partly due to the
much larger mass fraction of stars in galaxies than in
clusters. This also can lead to a significant increase of
the diffusion efficiency in such objects.

The low observed metallicity of the ISM \citep{Su2013} remains an important
issue in understanding the history of the star formation and evolution of early-type
galaxies.
Since the red giant winds,
planetary nebulae, and supernova explosions
serve as the main suppliers of the ISM gas, the
heavy-element abundances in the gas are expected
to be at least solar. Apart from its direct influence
on the heavy-element distribution, diffusion may
lead to considerable changes in the helium abundance profile.
Therefore, the assumption about a solar helium
abundance in analyzing the X-ray spectra can lead to
a significant bias in the measured metal abundance
\citep{Ettori2006,Markevitch2007,Medvedev2014}. 
Interestingly, an enhanced helium abundance
compared to the cosmological value could result
from diffusion already at the galaxy formation
stage, but this effect could not exceed a few tenths
of a percent \citep{Medvedev2016}.

In this paper we calculate the diffusion of elements
in the ISM gas based on \Chandra\  and \XMM\ observational data.
We consider an idealized problem: without magnetic fields
and a deviation of the ISM from hydrostatic
equilibrium, and with a constant (in time) temperature
profile. The goal of this exercise was to 
estimate the role of diffusion among other physical processes
in a hot ISM plasma. By solving the full set
of Burgers equations, we demonstrate a nontrivial
dependence of the integrated effect on galaxy mass
and surrounding environment.

\section{The sample of galaxies}
\label{sec:galsample}
\begin{table*}
\begin{center}
\caption{
The early-type galaxies selected for diffusion calculations. The morphological type was taken from NED\protect\footnote{http://ned.ipac.caltech.edu/}, $D$ is the distance to the galaxy from \protect\citet{Tonry2001} 
(the method of surface brightness fluctuations, SB1200F),  $\re$\ is the effective radius from the RC3 catalogue \protect\citep{Vaucouleurs1991}, $L_X$ is the deprojected luminosity of the thermal emission within $4\,\re$  in the soft X-ray range 0.3--2 keV from \protect\citet{Nagino2009}
 recalculated for the distances from column 3, $\Mism$ is the mass of the interstellar gas within $4\,\re$, $\Mtot$ is the
total mass within $10\,\re$ found from Eq.~\ref{eq:M}, classification of galaxy environments  is based on  \protect\citep{Faber1989}: 0 for galaxies in the field, 1 for galaxies in groups and clusters, and 2 for galaxies at group center.}
\label{tab:sample}
\begin{tabular}{lccccccc}
\hline
\hline
Galaxy & Morphological& $D$, & $\re$, & $L_X$, & $\Mism$, & $\Mtot$, & Environment \\
    &   type  & Mpc & kpc & $10^{40}$ \ergs & $10^8\, \msun$  & $10^{11}\, \msun$  &      \\
\hline
IC 1459 & E3--4 & 29.2 & 4.9 & 2.1 & 9.6 & 14.0 & 1 \\
NGC 720 & E5 & 27.7 & 4.8 & 4.4 & 11.0 & 12.0 & 0\\
NGC 1316 & SAB0 & 21.5 & 8.4 & 5.7 & 27.0 & 29.7 & 1  \\
NGC 1332 & S0 & 22.9 & 3.1 & 1.8 & 2.7 & 7.4  & 2 \\
NGC 1395 & E2 & 24.1 & 5.7 & 3.2 & 11.4 & 19.2 & 1 \\
NGC 1399 & E1 & 20.0 & 3.9 & 13.5 & 17.7 & 14.2 & 2 \\
NGC 3923 & E4--5 & 22.9 & 5.5 & 3.8 & 11.7 & 11.7 & 1  \\
NGC 4472 & E2 & 16.3 & 8.2 & 21.4 & 58.0 & 31.9  & 2\\
NGC 4552 & E0--1 & 15.3 & 2.2 & 3.3 & 2.4 & 6.3 & 1 \\
NGC 4636 & E0--1 & 14.7 & 6.3 & 20.3 & 25.9 & 27.8  & 2 \\
NGC 4649 & E2 & 16.8 & 5.6 & 9.9 & 17.0 & 21.9 & 1 \\
NGC 5044 & E0 & 31.2 & 8.1 & 166.1 & 164.7 & 45.5 & 2 \\
NGC 5846 & E0--1 & 24.9 & 7.6 & 37.9 & 70.0 & 92.4  & 2 \\
\hline
\hline
\end{tabular}
\small
\\
\end{center}
\end{table*}

To investigate the diffusion, we use \Ngal\  galaxies
from \citet[hereafter NM09]{Nagino2009}. Their basic characteristics are presented
in Table~\ref{tab:sample}. These are all early-type galaxies and
have been studied in detail in the soft X-ray range
with the \ROSAT,\ \XMM, and \Chandra\  space
observatories \citep[NM09]{Humphrey2006, Fukazawa2006, Su2013}. For this
work we selected only those galaxies for which the
total number of counts with \XMM\ 
 (the MOS and PN detectors) within
four effective radii ($\re$) exceeded 12\,000. These
data have good statistics, which allows the spectrum
deprojection procedure to be used more reliably to
determine the 3D temperature and density profiles of
the ISM (for a detailed discussion of
the deprojection procedure, see NM09 and \citealt{Humphrey2011}). The selected \Ngal\ galaxies satisfy well
our main criterion: to consider galaxies with a wide
range of luminosities, various surrounding environment, and a
relatively regular X-ray morphology.

In the soft X-ray range 0.3--2 keV the galaxies
of our sample span two orders of magnitude in luminosity:
$L_X \sim 1$--$300 \times 10^{40}$ \ergs. The selected
galaxies can be separated by the environments type into
three groups: isolated (0), groups/clusters galaxies (1), and
galaxies located at group centers (2). A characteristic
feature of the galaxies in rarefied environments (0
and some galaxies 1) is a temperature profile with a
negative gradient or a nearly isothermal profile. These
galaxies typically have a smaller size and a mass
$\lesssim 10^{12}\ \msun$. In contrast, the galaxies located 
in relatively high-density environments (mostly 2) have
a positive temperature gradient, while their temperature distribution is
similar to the temperature profile of the ICM
gas in clusters \citep{Vikhlinin2006}. Studying and
comparing the diffusion in galaxies surrounded by different 
environments is of great interest, because the
diffusion rate and the structure of the diffusion flows
depend most strongly on plasma temperature. As has
been shown previously, sharp temperature gradients
in cool-core clusters change significantly the pattern
of gravitational sedimentation of all elements 
\citep{Chuzhoy2004,Peng2009,Shtykovskiy2010,Medvedev2014}. 
The brightest galaxy from our
sample, NGC~5044, is the brightest cluster galaxy
(BCG). The recent observations with \GALEX\  
show strong ultraviolet excess in the spectrum of such objects (also known as the UV upturn)
 whose origin is still unknown \citep[see][]{Ree2007}.
Studying the diffusion in BCGs is very interesting, because the
UV upturn can be associated with an extremely high
helium abundance in the ISM \citep{Peng2009a}.

A regular X-ray morphology and the absence of
a significant asymmetry and large-amplitude surface
brightness perturbations suggest dynamical relaxation
of the system and hydrostatic equilibrium of the
gas in the galactic gravitational potential. Almost
all of the galaxies we investigate have regular X-ray
morphologies. The hydrostatic equilibrium condition
for the gas in the galaxies NGC~720, NGC~4472,
and NGC~4636 is discussed in detail in \citet{Humphrey2006,Humphrey2011}; for a discussion of this condition in
the remaining galaxies of our sample, see \citet[NM09]{Fukazawa2006}. The galaxy NGC~4636,
in which large-scale X-ray brightness perturbations
were detected \citep{Jones2002}, is an exception.
Two galaxies, NGC~1316 and NGC~1332, are lenticular.
The diffuse component of the X-ray emission
from NGC~1332 was studied in detail by \citet{Humphrey2004}. The galaxy NGC~1316 is a merger remnant
in the Fornax cluster; this galaxy was studied in
detail in X-rays by \citet{Kim2003}.

\section{Analysis of X-ray data}
\label{sec:profiles}
We use the \Chandra\ and \XMM X-ray
spectroscopy. The primary data reduction and the
determination of the ISM temperature and
luminosity were made in NM09. We use the luminosities
and temperatures from Table~4 in NM09,
which were estimated through a spectral analysis of
the deprojected spectra accumulated in annuli with
radii up to $8\,\re$. The maximum radius for which the
data are available for all galaxies of this sample is $4\,\re$.
In NM09 the spectra were deprojected by the ``onion-peeling''
method; one can familiarize oneself with its
algorithm, for example, in \citet{Buote2000}. NM09 used
the \Chandra\ and \XMM\  data for the central
region (1--2 $\re$) and the region $\leq 8 \re$, respectively.
The ISM gas temperature was determined by
fitting the spectra by the sum of \texttt{vAPEC} \citep{Smith2001} and a power-law model with a photon
index of 1.6 corresponding to the contribution of low-mass
X-ray binaries. We recalculate the luminosities
derived in NM09 for the distances to the galaxies
from \citet{Tonry2001}.

The luminosity of a spherical shell is defined as
\begin{equation}
L_i = \int\limits_{R_i}^{R_{i+1}}\Lambda_{A_i}(T_i) n_e^2(r) X \, dV = \Lambda_{A_i}(T_i) \int\limits_{R_i}^{R_{i+1}} n_e^2(r) X \, dV ,
\label{eq:L}
\end{equation}
where $L_i,A_i$ and $T_i$ are the luminosity, element
abundance, and temperature in the spherical shell
with an inner radius $R_i$ and an outer radius $R_{i+1}$,
$dV = 4 \pi r^2 dr$, $X = n_i/n_e$ is the ratio of the ion
number density  ($n_i$) to the electron number density
($n_e$),  $X \approx 0.91$ for a solar gas composition. The
integrated emissivity $\Lambda_{A_i} (T_i) = \int \epsilon(E,T_i)dE$
is calculated in the coronal approximation for a hot
optically thin plasma with a specific value of element abundance
$A_i$ (\texttt{AtomDB/APEC}, version 3.0.6, \citealt{Foster2012}).
We use the values for the oxygen-group (O, Ne, Mg), silicon-group
(Si, S), and iron-group (Fe, Ni) elements found
in NM09; the abundances of the remaining elements
were assumed to be solar. The solar abundance is
specified in accordance with NM09 from \citet{Anders1989}.
 For most galaxies the abundance
is specified by a single value for all radii: $A_i = A$; for
NGC~4472, NGC~4636, NGC~4639, and NGC~4649
we use the values in each spherical shell.

The electron number density $n_e$ for the galaxies of
the sample is most often described well by a simple
beta profile:
\begin{equation}
n_e = n_{e0} (1 + (r/r_c)^2)^{-3\beta/2}.
\label{eq:n_e1}
\end{equation}
For some of the bright galaxies, following NM09, we
use a modified beta profile whose slope changes at
radius $r_{c2}$:
\begin{equation}
n_e = n_{e0} [1 + (r/r_c)^2]^{-3\beta/2} [1 + (r/r_{c2})^2]^{-3(\beta_2-\beta)/2}.
\label{eq:n_e2}
\end{equation}
For faint galaxies with a large central bin we fix the
core radius $r_c = 0.05\ \re$. As was shown in NM09
based on \Chandra\ data, $r_c$  is always close to this value
for the galaxies in the sample.

Using $L_i$, $A_i$, $T_i$, and Eq.~(\ref{eq:L}), we determine the
parameters of the density profile; the values found are
given in Table~\ref{tab:best-fit}. The large value of $\beta$ for NGC~4552
is probably related to the small region of available
data $\leq4\,\re$. The data were obtained within $6\,\re$
for NGC~4649 and within $8\,\re$ for all the remaining
galaxies.

\begin{table*}
\begin{center}
\caption{Parameters of the ISM density and temperature profiles.
$a$ and $b$ --- Eq.~(\ref{eq:n_e1})~and~(\ref{eq:n_e2}) for the density, respectively;
$c, d$ and $e$ --- Eq.~(\ref{eq:T1}),~(\ref{eq:T2})~and~(\ref{eq:T3}) for the
temperature, respectively; $*$ --- the parameter is fixed during the fit.
}
\label{tab:best-fit}
\begin{tabular}{lccccccccc}
\hline
\hline
Name &$n_0$,& $r_c$,& $\beta$ & $r_{c2},$ & $\beta_2$ & $T_0$, & $r_t$, & $a$ & $T_{{\rm bg}}$/$T_{{\rm ICM}}$, \\
      & $10^{-1}\mbox{cm}^{-3}$ &  $\re$  &   &  $\re$ & & keV & $\re$   &   & keV     \\
\hline
IC $1459^{a,c}$  & 1.63 & $0.05^{*}$ & 0.43 & -- & --  & 0.62 & -- & -- & --  \\
NGC $720^{a,d}$  & 0.19 & 0.35 & 0.46 & -- & -- & 0.61 & $4.00^{*}$ & 0.21 & 0/--\\
NGC $1316^{a,c}$ & 1.33 & $0.05^{*}$ & 0.46 & -- & -- & 0.71 & -- & -- & --\\
NGC $1332^{a,d}$ & 3.04 & $0.05^{*}$ & 0.48 & -- & -- & 0.16 & $2.00^{*}$ & 2.00 & 0.46/-- \\
NGC $1395^{a,e}$ & 0.14 & 0.26 & 0.41 & -- & --  & 0.58 & 1.573 & 2.86 & --/0.77 \\
NGC $1399^{b,e}$ & 4.64 & $0.05^{*}$ & 0.46 & 1.09 & 0.25 & 0.83 & 1.79 & 1.68 & --/1.45 \\
NGC $3923^{a,d}$ & 1.14 & 0.10 & 0.50 & -- & -- & 0.63 & 0.32 & 0.08 & 0/-- \\
NGC $4472^{b,d}$ & 1.59 & $0.05^{*}$ & 0.48 & 0.46 & 0.30 & 0.62 & 7.90  & 1.5 & --/1.69 \\
NGC $4552^{a,d}$ & 1.10 & 0.37 & 0.63 & -- & -- & 0.35 & 0.74 & 0.78 & 0.41/-- \\
NGC $4636^{b,e}$ & 1.62 & $0.05^{*}$ & 0.20 & 0.16 & 0.50 & 0.51 & 0.92 & 2.08 & --/0.83 \\
NGC $4649^{a,e}$ & 2.88 & 0.03 & 0.40 & -- & -- & 0.88 & 0.11 & 1.29 & --/0.96 \\
NGC $5044^{b,e}$ & 0.53 & 0.13 & 0.25 & 2.20 & 0.44 & 0.70 & 4.07  & 1.74 & --/1.52 \\
NGC $5846^{b,e}$ & 1.29 & $0.05^{*}$ & 0.31 & 4.12 & 1.13 & 0.62  & 7.90 & 1.51 & --/1.69 \\
\hline
\hline
\end{tabular}
\end{center}
\end{table*}

The measured radial temperature profiles of the
ISM cannot be described by a universal
model. We use three simple models to describe
the temperature profile in different galaxies. The first
model is isothermal:
\begin{equation}
T = T_0 = {\rm const}.
\label{eq:T1}
\end{equation}
The next model is a temperature profile with a negative
gradient:
\begin{equation}
T = T_0 (1 + (r/r_t)^{2})^{-a} + T_{{\rm bg}}.
\label{eq:T2}
\end{equation}
Finally, the galaxies with a positive gradient are
described by a function that is commonly used to describe
the cool cores of galaxy clusters \citep[see][]{Vikhlinin2006}:
\begin{equation}
T =  T_{{\rm ICM}}\frac{(r/r_t)^{a} + T_0/T_{{\rm ICM}} }{1 + (r/r_t)^{a}}.
\label{eq:T3}
\end{equation}
The best-fit parameters of the temperature profile are
presented in Table~\ref{tab:best-fit}.

Assuming a hydrostatic equilibrium of the ISM
 and a spherically symmetric distribution,
we determine the galaxy mass profile $M(<r)$:
\begin{equation}
M(<r) = \frac{k_b T r}{G \mu m_p} \left( \frac{\log{n}}{\log{r}} + \frac{\log{T}}{\log{r}} \right),
\label{eq:M}
\end{equation}
where $k_b$ and $G$ are the Boltzmann and gravitational
constants, respectively; $\mu$ is the mean molecular
weight, $\mu \approx 0.62$ for solar element abundances;
and $n = (1 + X) n_e$ is the particle number density.
The derived temperature and density profiles of the
ISM as well as the mass profiles are
presented below in Fig.~6.

\section{Diffusion calculations}
\label{method}
Depending on the strength and topology of the
magnetic field as well as the MHD turbulence characteristics,
the transport processes in the ISM can
be significantly suppressed by the magnetic fields. 
A simple  approach is to introduce a constant
suppression factor \citep[in the context of diffusion in the ICM,
see][]{Peng2009}: $S_B \equiv \kappa / \kappa_{sp} \leqslant 1$, 
where $\kappa_{sp}$ is the Spitzer thermal conductivity of
a non-magnetized plasma \citep{Spitzer1962}. 
The current theoretical constrains on the suppression magnitude 
is  provided by numerical simulations only for the ICM plasma in galaxy clusters.
In this case  the global thermal
conductivity is expected to be moderately suppressed 
 in a tangled magnetic field of the ICM plasma: 
$S_B \sim 0.1$--$0.2$  \citep{Chandran1998, Narayan2001, Chandran2004}. 
Besides that, the weakly collisional magnetized plasma of the ICM
is subject to a wide variety of instabilities, such as 
the firehose and mirror instabilities, which 
 can suppress transport efficiency by another factor of 5--10
\citep{Riquelme2016,Komarov2016}.
%to the firehose and mirror instabilities that suppress
%the thermal conductivity approximately by a
%factor of 5:  $S_B \sim 0.2$ . The
%generation of whistlers is possible in the case of a
%significant electron pressure anisotropy, which also
%leads to a suppression of the thermal conductivity
%by a factor of 3--4, $S_B \sim 0.3$ \citep{Riquelme2016}. 
%These estimates were obtained for the
%intergalactic gas of galaxy clusters.
However the collision mean free path and other physical properties of the ISM can be considerably
different from that of the ICM, therefore, the role of the above-mentioned
instabilities remains unclear and requires further numerical simulations. In this paper the ISM
plasma is assumed to be non-magnetized ($S_B =1$), i.e., we estimate the maximum possible effect of diffusion.

To calculate the element diffusion in the ISM, 
we use the same numerical scheme as developed by
\citet{Shtykovskiy2010} to model the diffusion
in the ICM. % and described below
We consider the Burgers' flow equations for the mass,
momentum, and energy conservation.
Under the assumption of hydrostatic equilibrium, spherical symmetry and
the same temperature for all gas species the Burgers' equations are \citep{Burgers1969,Thoul1994}
\begin{equation}
\frac{\partial{n_s}}{\partial{t}}+\frac{1}{r^2}\frac{\partial}{\partial{r}}[r^2n_s\,(w_s+u)]=0,
\label{eq:continuity}
\end{equation}
\begin{multline}
\frac{d(n_sk_BT)}{dr}+n_sm_sg-n_sZ_seE=\sum_{t\neq
  s}K_{st}[(w_t-w_s)+\\
+0.6(x_{st}r_s-y_{st}r_t)],
\label{eq:momentum}
\end{multline}
\begin{multline}
\frac{5}{2}n_sk_B\frac{dT}{dr}=\sum_{t\neq
  s}K_{st}\{\frac{3}{2}x_{st}(w_s-w_t)-\\
-y_{st}[1.6x_{st}(r_s+r_t)+Y_{st}r_s-4.3x_{st}r_t]\}-0.8K_{ss}r_s,
\label{eq:energy}
\end{multline}
where $g=[GM(r)/r^2]$ is the gravitational
acceleration modulus, $E$ is the radial induced
electric field, $m_s$ is the mass of particles of species $s$,
$\mu_{st}=m_s m_t/(m_s+m_t)$, $x_{st}=\mu_{st}/m_s$, $y_{st}=\mu_{st}/m_t$ and
$Y_{st}=3y_{st}+1.3x_{st}m_t/m_s$. The numerical coefficients
in these equations correspond to Coulomb
potential with a long-range cut-off at the Debye length. The gas viscosity
in Eqs.~\ref{eq:momentum}--\ref{eq:energy} is assumed to be negligible.

The friction coefficient $K_{s,t}$ between particles of
species $s$ and $t$ for a Coulomb interaction potential
is defined by the expression \citep{Chapman1991}:
\begin{equation}
K_{st}=(2/3)\mu_{st}(2k_BT/\mu_{st})^{1/2}n_sn_t\sigma_{st},
\label{eq:Kst}
\end{equation}
where the transport cross-section is:
\begin{equation}
\sigma_{st}=2\sqrt{\pi}e^4Z_s^2Z_t^2(k_BT)^{-2}\ln\Lambda_{st}.
\label{eq:cross_section}
\end{equation}

Since the ISM gas is considerably denser
and colder than the ICM, the Coulomb
logarithm, $\ln\Lambda_{st}$, can be considerably smaller than 40 (the typical value for the ICM conditions).
Nevertheless, the condition $\Lambda_{st} \gg 1$ (ideal plasma) is
always fulfilled for the ISM plasma. Therefore,
we use the approximate formulas by \citet{Huba2009}
for a Coulomb potential with shielding at the Debye
length. Denote the gas temperature in eV by  $T_{{\rm eV}}$ and
the plasma electron component by the subscript $e$.
For the ion--ion interaction
\begin{equation}
\ln\Lambda_{st} =  23 - \ln \left[ \frac{Z_s Z_t}{T_{{\rm eV}}^{3/2}} (n_s Z_s^2 + n_t Z_t^2)^{1/2}\right].
\label{eq:coulomb1}
\end{equation}
For the electron--electron interaction
\begin{equation}
\ln\Lambda_{ee} = 23.5 - \ln n_e^{1/2} T_{{\rm eV}}^{-5/4} - [10^{-5} + (\ln T_{{\rm eV}} - 2)^2 / 16]^{1/2}.
\label{eq:coulomb2}
\end{equation}
For the ion--electron interaction, if $T_{{\rm eV}} m_e/m_s < T_{{\rm eV}} < 10 Z_s^2\, {{\rm eV}}$,
\begin{equation}
\ln\Lambda_{se} = 23 - \ln n_e^{1/2} Z_s T_{{\rm eV}}^{-3/2},
\label{eq:coulomb3}
\end{equation}
and for $T_{{\rm eV}} m_e/m_s <  10 Z_s^2\, {{\rm eV}} < T_{{\rm eV}} $:
\begin{equation}
\ln\Lambda_{se} = 24 - \ln n_e^{1/2} T_{\rm eV}^{-1}.
\label{eq:coulomb4}
\end{equation}
For all interactions $\ln\Lambda_{st} = \ln\Lambda_{ts}$. In the ranges
of temperatures and densities of interest to us the
Coulomb logarithm lies within the range 30--37.
When passing from Eq.~(\ref{eq:coulomb3})~to~(\ref{eq:coulomb4}), a discontinuity
emerges in the derivative of the diffusion velocity.
To prevent the ensuing numerical instability of our
calculations, we used Eq.~(\ref{eq:coulomb3}) for ions with a nuclear
charge $>+8$ and Eq.~(\ref{eq:coulomb4}) for ions with a smaller
charge. The values of the Coulomb logarithm obtained
in this way lie within the accuracy of the
approximate formulas, 10\% \citep{Huba2009}.

The thermal diffusion occurs due to the interference between  Eq.~\ref{eq:momentum} and~\ref{eq:energy}
 via the quantities $r_s$, which are the so-called residual heat flow velocities 
defined as the local heat flow transported by particles of species $s$,
divided by the partial pressure $p_s$ \citep{Burgers1969}.
A phenomenological explanation of the thermal diffusion can be found, e.g., in
\citet{Monchick1967}. For the the Coulomb cross-section given by Eq.\ref{eq:cross_section},
the thermal diffusion tends to move the more highly charged and 
more massive particles up the temperature gradient. 
%It is because a test heavy ion interacts more with low energy
%field particles. In the presence of a temperature gradient,
%tgave on average larger energy than those g
Note that
even in the absence of a temperature gradient the
total local heat flow $Q = \sum r_s n_s k_b T$, even if small, is
nonzero  \citep{Chuzhoy2004}. As in the case of
galaxy clusters, we deemed the temperature profile to
be constant in time ($T(t) = {\rm const}$) during the entire
calculations by assuming an equilibrium between the
heating and cooling of the ISM.

The diffusion velocities obey the mass and charge
(the absence of a current) conservation laws that
should be added to the set of equations~(\ref{eq:momentum})--(\ref{eq:energy}):
\begin{equation}
\sum \rho_iw_i=0,
\label{eq:flows}
\end{equation}
\begin{equation}
\sum Z_in_i w_i=0.
\label{eq:currents}
\end{equation}

As a result of sedimentation, the mean molecular
weight of particles in the galactic central region,
$\mu = \sum m_i  n_i / m_p n$, increases. This causes the total local
gas pressure to decrease and the hydrostatic equilibrium
to be upset. However, the system rapidly returns
to an equilibrium state due to the net mass flow into
the galactic central region:
\begin{equation}
\frac{du}{dt}=-\frac{\nabla p}{\rho}-g.
\label{eq:euler2}
\end{equation}

Since the characteristic diffusion time scale is
always larger than the characteristic hydrodynamic
time by many orders of magnitude, Eqs.~(\ref{eq:momentum})--(\ref{eq:energy})
remain applicable, because the inertial terms ($\sim du_s/dt$) in Eqs.~(\ref{eq:momentum})--(\ref{eq:energy}) make a negligible
contribution. Equations~(\ref{eq:continuity})--(\ref{eq:euler2}) completely define
the diffusion velocities  $w_s$, the quantities $r_s$, the low
gas flow velocity $u$, and the induced electric field
strength $E$.

\subsection{Description of the Model}
To solve the set of equations~(\ref{eq:continuity})--(\ref{eq:euler2}), we use a
homogeneous spatial grid covering the region from 0
to $10\,\re$ (22--85 kpc) for each galaxy. The spatial
grid step is 7--28 pc. The time step is specified
so as to ensure the numerical stability of the scheme,
$dt =2.5\times 10^3$ yr. We limit the diffusion calculation
by 2 Gyr.

As an outer boundary condition for the continuity
equation, the particle number density was specified
to be equal to the initial one at the outer boundary of
the computational domain. This boundary condition
qualitatively corresponds to a galaxy placed in an infinite
gas reservoir with a fixed density. To calculate the
velocities (Eqs.~(\ref{eq:momentum}), (\ref{eq:energy}), (\ref{eq:euler2})) at the outer boundary,
we used one-sided derivatives. At $r=0$ the number
density was calculated by expanding Eq.~(\ref{eq:continuity}) near zero
\citep[see][]{Shtykovskiy2010}. The velocities
at $r=0$ were assumed to be zero.

The derived analytical expressions (Eq.~\ref{eq:n_e1}--\ref{eq:T3}) for the gas temperature
and density were extrapolated to $10\,\re$. This
reduces the influence of outer boundary conditions on
the results of our calculations for the region of galaxies
under study ($r\leq 8\, \re$). We tested our solution
by replacing the outer boundary condition with an
``opaque wall'' (all velocities are zero). For the galaxy
with the strongest diffusion at large radii, NGC~4636,
this replacement leads to significant changes only for
the region $r>9\,\re$ ($\sim 10$\% in the change of He to
H). For the mass of any element within $r<8\,\re$ the
difference in 2~Gyr does not exceed 8\%; the difference
within $r<4\,\re$ is less than 4\%. The net mass flow
(Eq.~(\ref{eq:euler2})) causes the gas mass in the galaxy to increase
by 5\% in 2~Gyr of diffusion. For the remaining
galaxies the differences turn out to be even smaller.

We also checked that the Coulomb mean free path
(Eq.~\ref{eq:cross_section}) is always much smaller than the size of the
computational region. In the temperature and density
ranges under consideration the mean free path lies
within the range $10^{-9}< \lambda_C < 0.1$ kpc.

The abundances of all elements at the initial time
of our simulations were assumed to be solar \citep{Lodders2003}
 and identical at all radii. Although the
heavy-element abundances in the interstellar gas of
early-type galaxies can be considerably lower than
the solar ones \citep{Su2013}, this affects weakly
the diffusion calculations, because the heavy elements
are a small admixture in the H--He plasma.

\begin{figure*}
\centering
\includegraphics[width=0.8\linewidth]{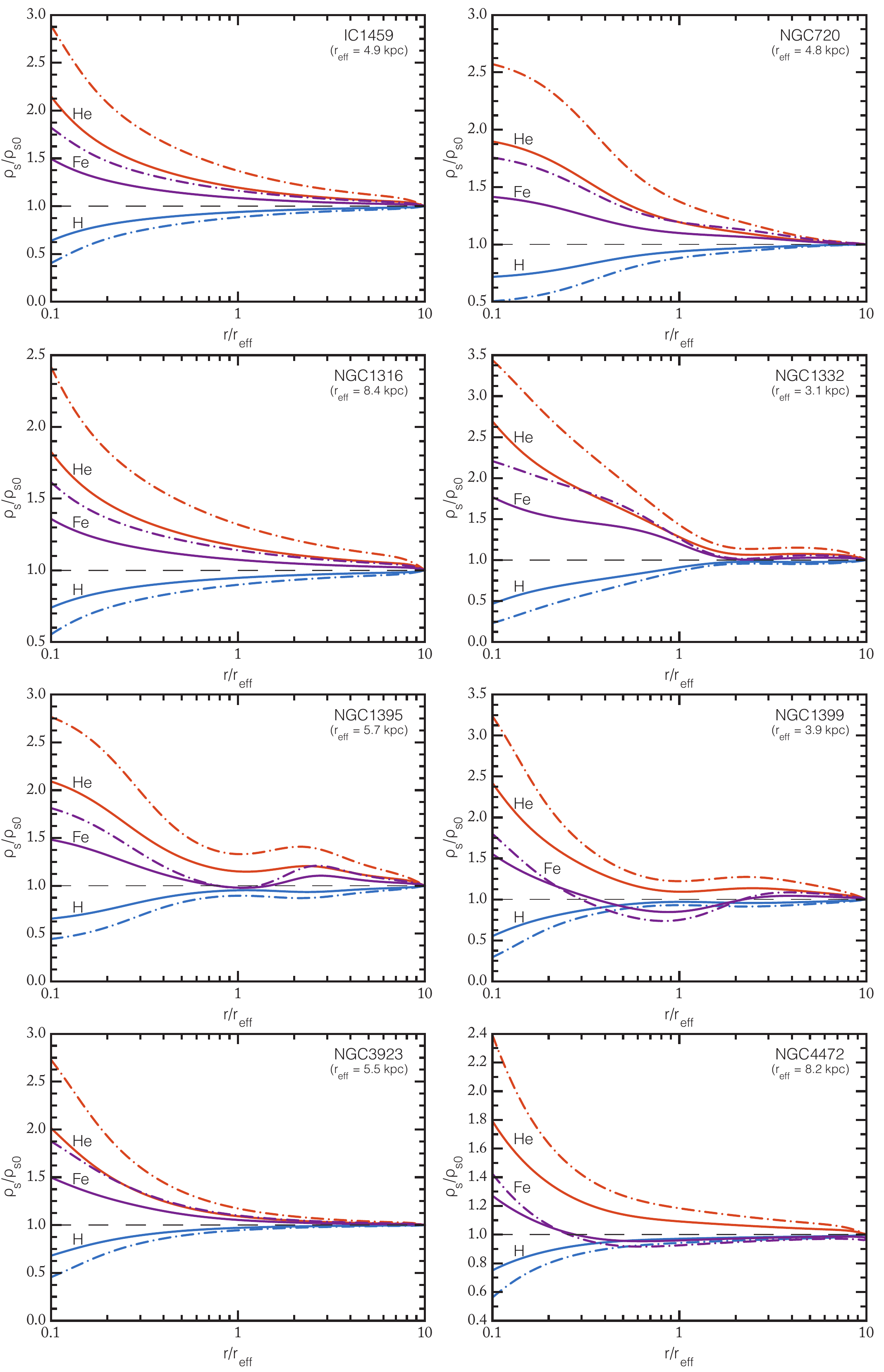}
\caption{Hydrogen (blue lines), \ion{He}{III} (red lines), and \ion{Fe}{XXII} (purple lines) density distributions as a function of radius (in
units of the effective radius) after 1 (solid lines) and 2 (dash-dotted lines)~Gyr of diffusion. The results of our calculations are
presented for the galaxies from Table~\ref{tab:sample}.}
\label{fig:results1}
\end{figure*}
\begin{figure*}
\centering
\hspace{1.63cm}
\includegraphics[width=0.8\linewidth]{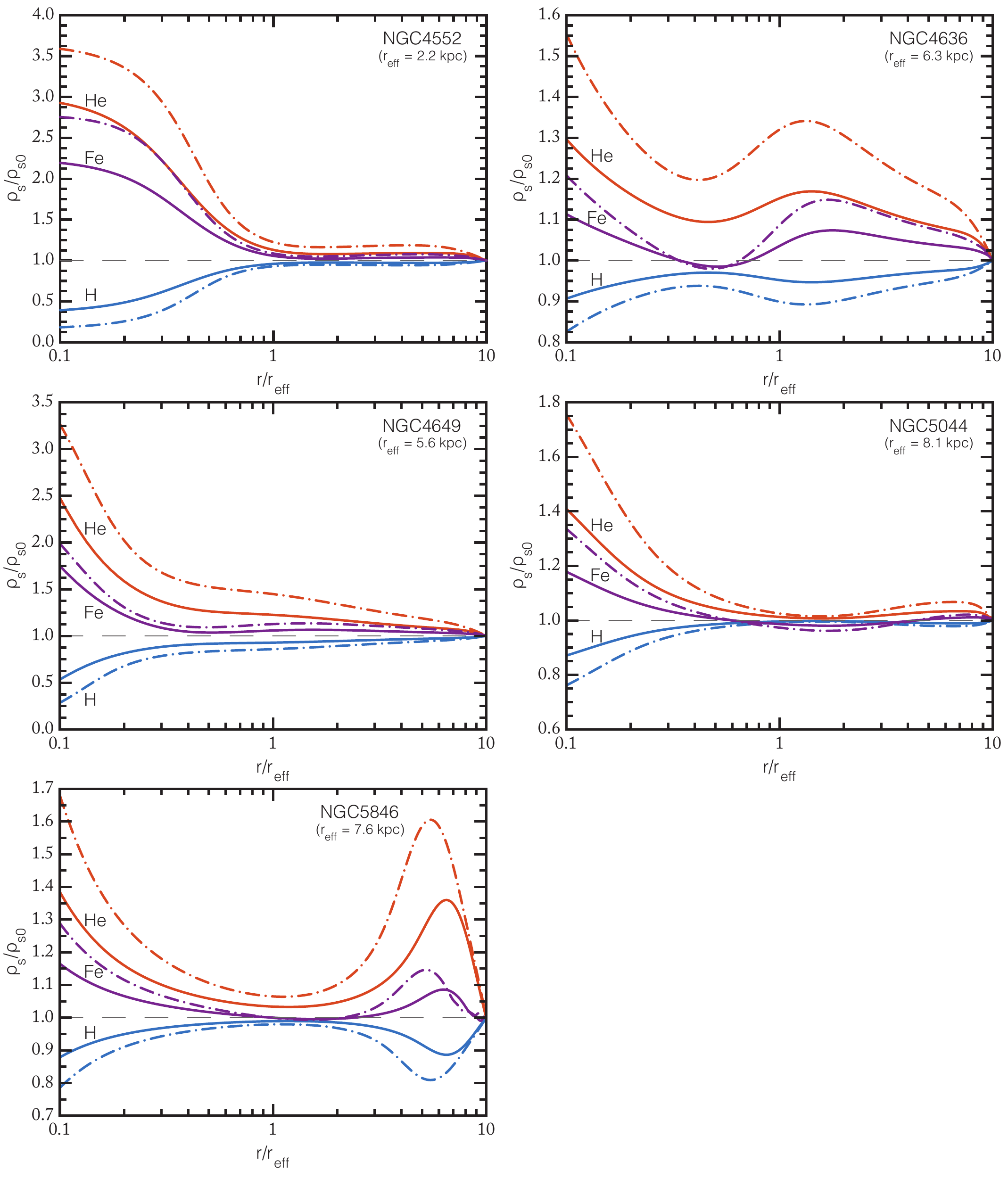}
\contcaption{}
\label{fig:results2}
\end{figure*}

\section{Results}
\label{results}
\subsection{The Spatial Distribution of Elements}
The spatial distributions of elements after 1 and
2~Gyr of diffusion are presented in Fig.~\ref{fig:results1}. The results
for hydrogen (blue lines), helium (red lines),
and \ion{Fe}{XXII} (purple lines) are shown. \ion{Fe}{XXII} was
chosen to demonstrate the diffusion of heavy ions; a
slight change in the ionization fraction ($\pm 2$) affects
weakly the result (within 1\%). For all of the chosen
galaxies diffusion leads to an increase in the abundances
of elements heavier than hydrogen within $\re$.
The greatest effect is expected for helium. A strong
dependence of the element abundance profile on temperature
gradient is seen in Fig.~\ref{fig:prof1}. The temperature
gradient exerts the greatest influence on the iron diffusion
due to the dependence of thermal diffusion on
nuclear charge. Thus, accurate measurements of the
heavy-element abundances in the ISM
could provide additional information both about the
gas temperature profile and about the transport efficiency
in the gas (or the magnetic field strength).
However, this method will probably be difficult to
apply in practice, because it is necessary to take into
account the enrichment of the interstellar medium
with heavy elements from stellar winds and during
supernova explosions as well as the turbulent mixing
and resonant scattering in metal lines \citep{Gilfanov1984,Zhuravleva2010}.

\subsection{Change in the Mass Fraction of Elements}
\begin{figure*}
\centering
\includegraphics[width=0.8\linewidth]{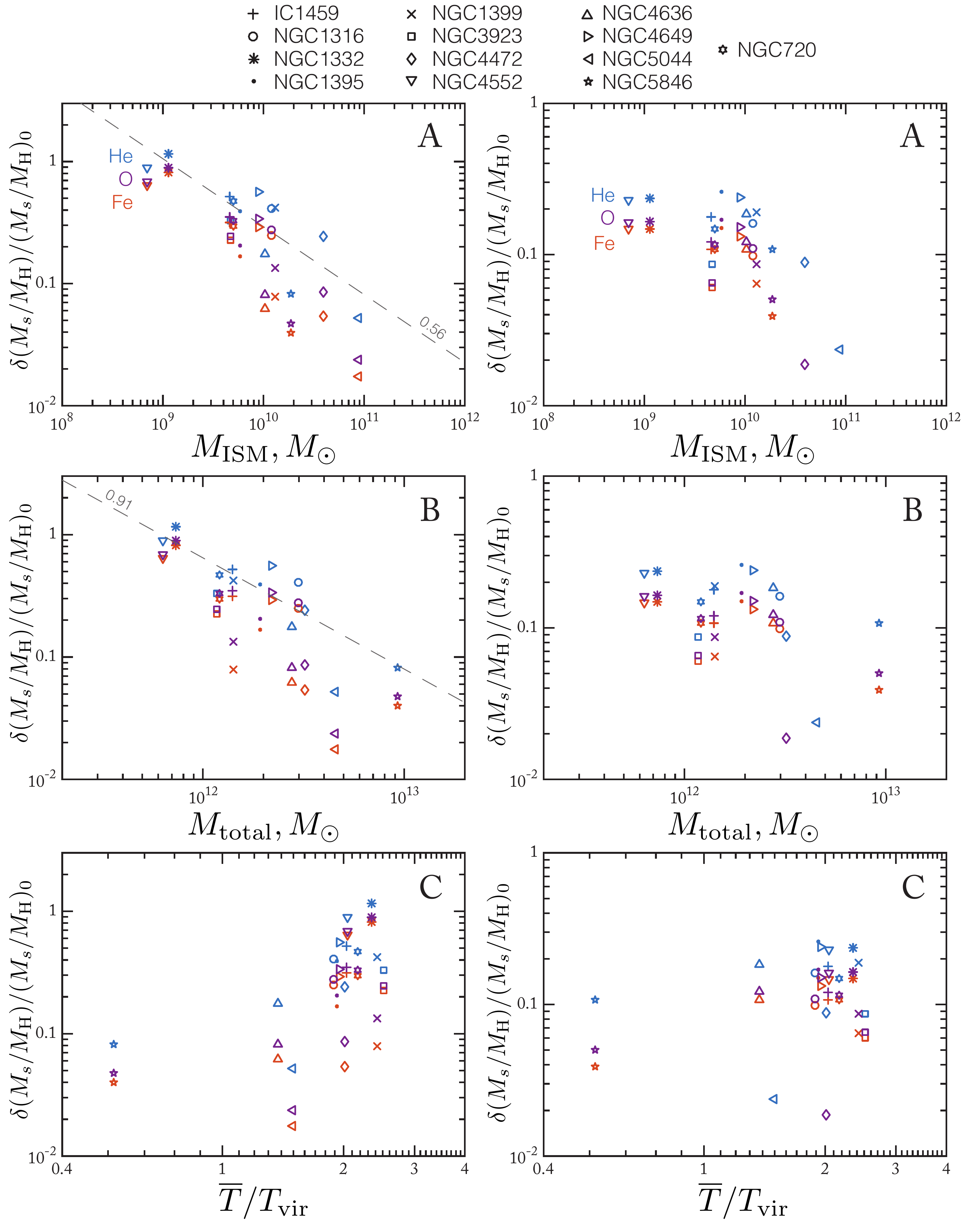}
\caption{Relative change in the  \ion{He}{III} (blue symbols),   \ion{O}{VIII} (purple symbols), and \ion{Fe}{XXII} (red symbols) mass divided by the hydrogen mass within 1 (left panels A, B, and C) and 4 (right panels A, B, and C) effective radii after 1~Gyr of diffusion as
a function of the interstellar gas mass within $10\, \re$ (panel A), the total mass within $10\, \re$  (B), and the mean temperature $\overline{T}$ (determined from Eq.~\ref{eq:T1}, panel C) divided by $T_{\rm vir} = \gamma \mu m_p G M / (3 \times 10\, \re)$ (for more details, see the text). The gray dashed line indicates the fit to the helium change by a power law with its index specified near the line.
}
\label{fig:dM}
\end{figure*}

\begin{figure*}
\centering
\hspace{2cm}
\includegraphics[width=0.7 \linewidth]{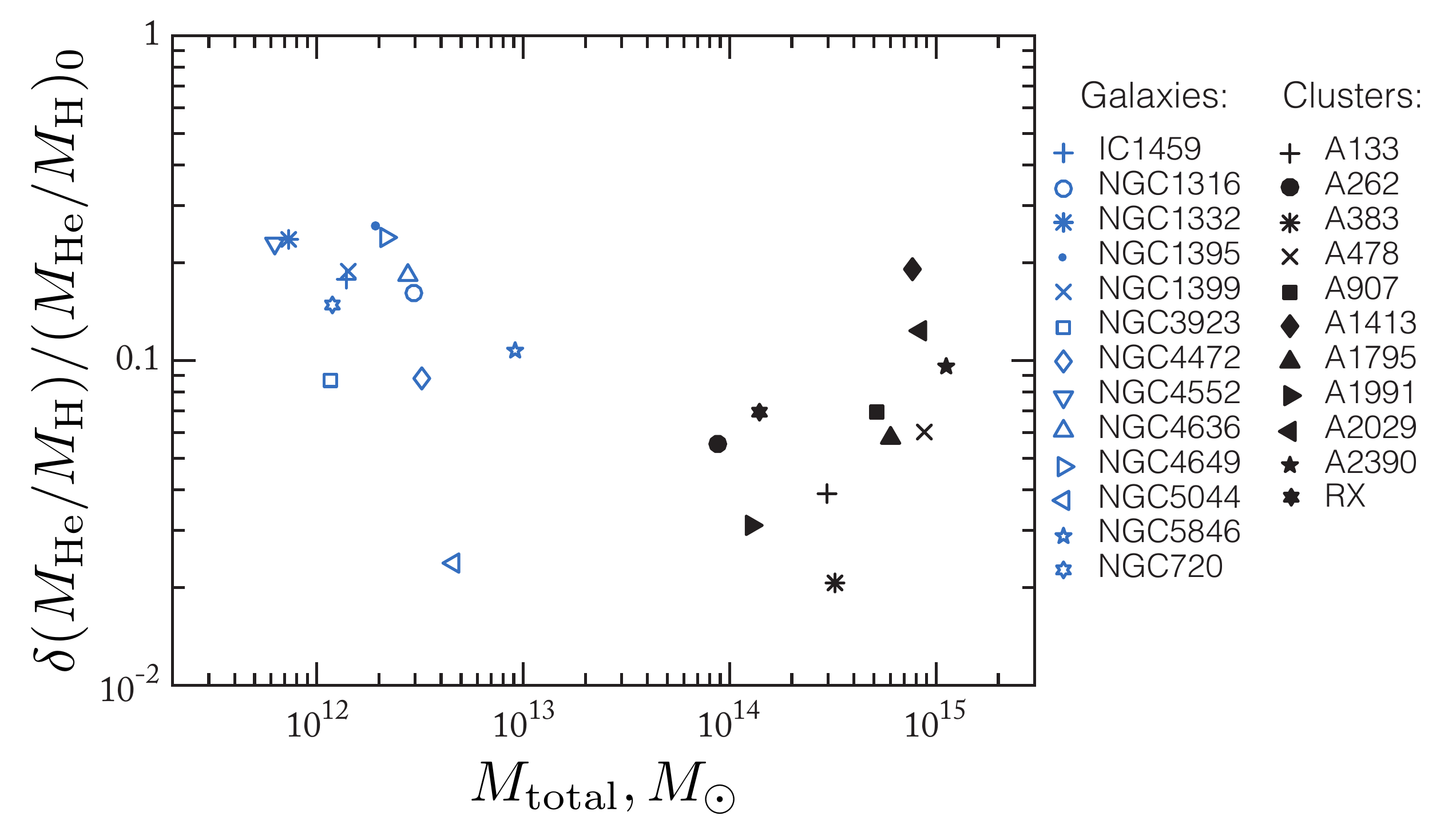}
\caption{Relative increase in the helium mass divided by the hydrogen mass for the galaxies of our sample and clusters from
\citet{Vikhlinin2006} as a function of the total mass after 1~Gyr of diffusion. We calculated the hydrogen and helium mass
for the galaxies within $4\,\re$ and the total mass within $10\, \re$. For the galaxy clusters the corresponding sizes were taken to be
$0.08\,r_{\rm 500} $ and $r_{\rm 500}$ (see the text). ``RX'' denotes the ``X-ray superbright elliptical galaxy'' RX~J1159$+$5531.
}
\label{fig:comp_cl}
\end{figure*}

We see that for most galaxies (NGC~1332,NGC~1395,
NGC~1399, NGC~4472, NGC~4636, NGC~5044,
NGC~5846) the effect of thermal diffusion dominates
over the gravitational sedimentation of iron in the
places where the temperature gradient is largest. All
these galaxies, except NGC~1395, are at the group
center (environment type~2, see Table~\ref{tab:sample}). The element
sedimentation profile in the galaxy NGC~4636, where
helium (and to a lesser degree iron) is accumulated
in the entire observed region ($\re$--$8\,\re$), is of greatest
interest. In the remaining cases, the greatest increase
in element abundances occurs in a small central
region, $r < 0.3\ \re \sim 0.7$--$2.5$ kpc. The huge effect
of diffusion at $r \sim 5$--$6\,\re$ for the galaxy NGC~5846 is
associated with a sharp drop in density at $r > 4\,\re$
and simultaneously a high gas temperature. For
comparison, diffusion in NGC~4552, on the whole,
affects weakly the element abundances, despite the
fact that the gas density at large radii in NGC~4552
and NGC~5846 is of the same order of magnitude.

The right and left panels in Fig.~\ref{fig:dM} show the change
in the mass contained in different elements in the
galaxies being studied after 1~Gyr of diffusion within
$\re$ and $4\,\re$, respectively. Apart from the iron diffusion,
we calculate the diffusion of the main groups
of heavy elements: \ion{O}{VIII} and \ion{Si}{XIII} (the result for
\ion{Si}{XIII} is not shown, because the points are close to
\ion{O}{VIII}, but listed in Table~\ref{tab:results}). The results are shown in the form of dependences
of the relative change in the mass of an element
divided by the hydrogen mass ($\frac{M_{s}}{M_{{\rm H}}} / (\frac{M_{s}}{M_{{\rm H}}})_{t=0} - 1$)
 on (A) the ISM gas mass within $10\, \re$ (i.e.,
the total volume of the computational domain) found
by integrating the gas density, (B) the total mass
within $10\, \re$ derived from the hydrostatic equilibrium
condition, and (C) the ratio $\overline{T}/T_{\rm vir}$, where $\overline{T}$
is the mean gas temperature obtained by fitting the
measured temperature profile by Eq.~(\ref{eq:T1}), and $T_{\rm   vir} = \gamma \mu m_p G M / (3 \times 10\, \re)$ 
(where $\mu =  0.62$, $\gamma = 1.2$) is
the ``virial'' temperature corresponding to the galactic
gravitational potential \citep{Makino1998}. The results
of our calculations are also presented in Table~\ref{tab:results}.

We see that the integrated effect of diffusion grows
with decreasing gas mass and total mass of the
galaxy. This is partly related to an increase in the ratio
$\overline{T}/T_{{\rm  vir}}$ for such objects (right panels). The decrease of
the gas fraction, $f_b$, in lower-mass galaxies serves as
an additional factor, with the diffusion velocity being $\sim f_b^{-1}$ \citep{Peng2009}. The increase in
the concentration of heavy elements, on the whole,
follows the change in the helium abundance.

\subsection{Comparison with the Diffusion in Galaxy Clusters}
It is interesting to compare the calculated change
in the abundances of elements due to diffusion
in early-type galaxies with the expected result of
diffusion in galaxy clusters; the sample of clusters
from \citet{Vikhlinin2006}  can be used in this
case. We use the results from \citet{Shtykovskiy2010} and \citet{Medvedev2014} for
the clusters A262 and A2029, respectively, and
calculate the diffusion in a similar way for the
remaining clusters using the best-fit formulas for the
temperature and density from \citet{Vikhlinin2006}
by extrapolating them to the cluster center. During
our calculations the ICM gas is assumed
to be in hydrostatic equilibrium in the gravitational
potential of a dark matter cluster halo with an NFW
density profile \citep{Navarro1997}. For a more
detailed description, see \citet{Shtykovskiy2010}.

To compare the results of diffusion in clusters and
galaxies, it is necessary to choose a spatial scale
within which the integrated effect could be calculated.
It is convenient to choose the radius $r_{\rm 500}$ as such
a scale. Unfortunately, the ISM region
accessible to measurements turns out to be smaller
than $r_{\rm 500}$ by an order of magnitude. Therefore, we
can give only a rough estimate of this radius for the
galaxies of our sample: $r_e \sim 0.02\,r_{\rm 500}$  (see Section~2).

Figure~\ref{fig:comp_cl} compares the relative increase of the
helium mass divided by the hydrogen mass for the
galaxies of our sample and clusters from \citet{Vikhlinin2006} as a function of the total mass after 1~Gyr
of diffusion. In the case of galaxies, we determine the
helium and hydrogen mass within $4\,\re$ and the total
mass within $10\, \re$. For the clusters we calculate the
helium and hydrogen mass within $0.08\,r_{\rm 500} $ and the
total mass within $r_{\rm 500}$; $r_{\rm 500}$ is taken from \citet{Vikhlinin2006}.
 The sedimentation amplitude in early-type
galaxies is seen to be comparable to that in
galaxy clusters. Interestingly, in the case of clusters
the effect is enhanced for more massive clusters, while
in the case of galaxies an inverse dependence is observed.
For the clusters this result is a consequence
of the increase in the temperature of the ICM with increasing cluster mass. We checked
that a change in the radius within which the effect
is calculated for galaxy clusters ($0.08\,r_{\rm 500} $) does not
affect the general dependence of the sedimentation
amplitude on cluster mass.

\subsection{Helium Sedimentation}
A significant increase of the helium abundance in
the entire ISM volume under consideration,
with the greatest value in the central region $<1\,\re$,
is expected from the results of our calculations. The
largest increase in the helium mass relative to hydrogen
is expected at the center of NGC~1332, 115\%
in 1~Gyr (by more than a factor of 2). This is one
of the least massive galaxies in our sample. For the
lowest-mass galaxy from the sample, NGC~4552, the
expected increase in the helium abundance within $\re$
is 88\%. A maximum increase in helium within $4\, \re$ is
expected for the large galaxy NGC~1395, 25.8\%, while
a maximum increase in the total helium mass (within
$10\,\re$) is expected for the galaxy NGC~5846, 26\%.

The deviation of the helium abundance from its
solar value turns out to be important in analyzing
the X-ray observations \citep{Drake1998,Ettori2006,Markevitch2007,Peng2009,Medvedev2014}.
 Since the helium
abundance cannot be directly estimated from X-ray
spectroscopy, the helium abundance is commonly
assumed to be solar when analyzing the spectra.
This assumption leads to a biased estimate of the
quantities derived from X-ray spectroscopy: the
gas temperature, density, and heavy-element abundances.
To find the amplitude of these biases, we perform
the following test \citep[see][]{Medvedev2014}.
Using the \texttt{vvAPEC} model (\texttt{AtomDB/APEC}, version
3.0.6, \citealt{Foster2012}) in the {\sc XSPEC} software
(heasoft v6.19, \citealt{Arnaud1996}), we generate a set of
thermal spectra for an optically thin hot plasma in
wide ranges of temperatures ($0.1 < T^{\rm tr} < 8$) and
helium abundances ($0 <  x^{\rm  tr} < 5$, in solar units). The
abundances of the remaining elements are assumed
to be solar \citep{Lodders2003}. Next, for each such
spectrum we model the spectrum that could be
measured with the \Chandra\ ACIS-S camera (the
data on the instrumental response are taken for
ACIS Cycle 18) by means of the {\it fakeit} command.
We assume that the data could be obtained with a
long exposure; therefore, no random noise is added
to the model spectra. As a result, the derived spectra
are fitted in the energy range 0.5--10 keV by the
\texttt{APEC} model, with the helium abundance having
been fixed at its solar value.

\begin{figure}
\centering
\includegraphics[width=1 \columnwidth]{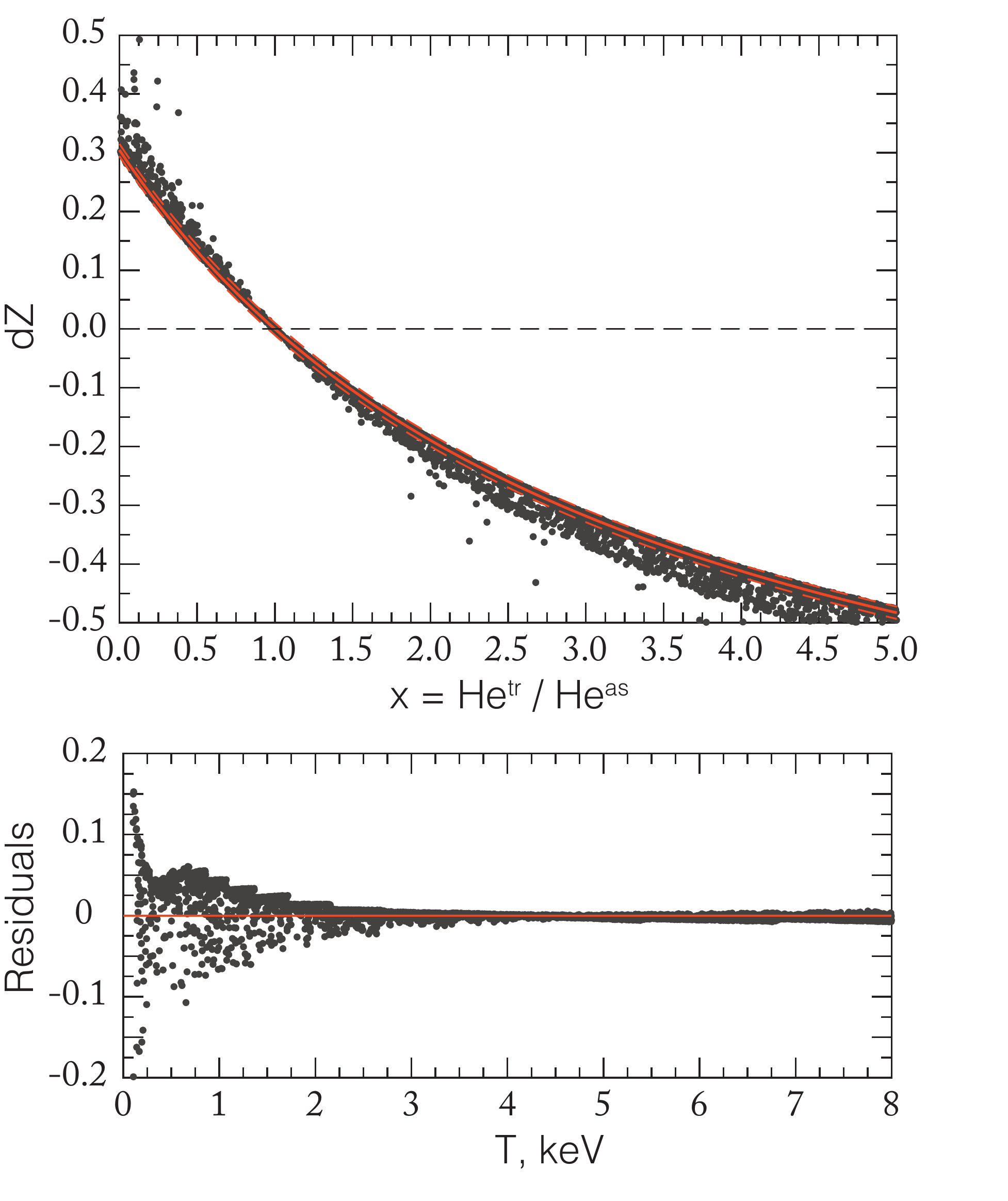}
\caption{Error in the metal abundance $dZ = Z^{{\rm tr}}/Z^{{\rm fit}} - 1$ under the erroneous assumption about the helium abundance
$x = {\rm He}^{{\rm tr}}/{\rm He}^{{\rm as}}$. The black dots correspond to the result of our numerical test (see the text), the solid line indicates the fit by
Eq.~(\ref{eq:approx}). The red dash-dotted line indicates the region that contains 95\% of the dots. The lower panel shows the residuals of
our numerical test and the fit as a function of gas temperature.}
\label{fig:helium_test}
\end{figure}

In the temperature range under consideration an
optically thin H--He plasma radiates mostly
by bremsstrahlung. In
that case, the integrated specific energy release depends
on the square of the nuclear charge of the ion
interacting with the electron. This leads to a change
of the continuum level in the plasma spectrum when
varying the helium abundance (${\rm He}^{{\rm tr}}$). Therefore, the
heavy-element abundances ($Z^{{\rm fit}}$) found by fitting this
spectrum by a model with a solar helium abundance
(${\rm He}^{{\rm as}}$) will be biased relative to those specified in
the model ($Z^{{\rm tr}}$). The shape of the bremsstrahlung
spectrum also depends on the nuclear charge, but to
a lesser degree (the charge dependence of the Gaunt
factor; see, e.g., \citealt{Rybicki1979}). This
effect leads to a biased estimate of the gas temperature
($T^{{\rm fit}}$).

The results of the test described above for $2\times10^4$
realizations are presented in Fig.~\ref{fig:helium_test}. There is a
significant bias of the heavy-element abundances that
can be described by a simple approximate formula:
\begin{equation}
dZ = (1 - x)/(3.28 + x),
\label{eq:approx}
\end{equation}
where  $x = {\rm He}^{{\rm tr}}/{\rm He}^{{\rm as}}$
is the deviation of the helium abundance from the solar one and
$dZ = Z^{{\rm fit}}/Z^{{\rm tr}} - 1$
is the error in the metal abundance.
The visible ``step''
on the panel with residuals is equal to the step in
the temperatures for which the spectroscopic \texttt{AtomDB}
data were computed and is associated with the {\sc XSPEC}
interpolation for intermediate temperatures. However,
this effect does not affect the main trend of
the residual. We made sure of this by applying the
\texttt{vMEKAL} and \texttt{MEKAL} models \citep{Kaastra1992}
in a similar way by calculating the thermal spectrum
for each temperature (without using any interpolation). The result obtained is analogous to the \texttt{APEC}
model. We see that the error in the metal abundance
increases for a temperature $T^{{\rm tr}}<4$ keV. Thus, at
$T^{{\rm tr}} \sim 0.1$ keV determining the heavy element abundances
by fitting the thermal spectrum at energies
$>0.5$ keV can lead to an uncertain result that will
depend strongly on the assumed helium abundance.
The associated error in the temperature determination turns out
to be insignificant, \mbox{$dT = T^{{\rm fit}}/T^{{\rm tr}}-1 < 0.03$}, with the
biased estimate of the metal abundance contributing
to this error.

A variation of the helium abundance ${\rm He}^{{\rm tr}}$ also leads
to a change in the ratio of the determined and specified
normalization quantities: $\eta \equiv \frac{N^{{\rm fit}}}{N^{{\rm tr}}} \sim \frac{n^{{\rm fit}}_H  n^{{\rm fit}}_e}{n^{{\rm tr}}_H n^{{\rm tr}}_e}$. However,
$\eta$ gives no direct information about the error in the gas
density. Fixing the solar hydrogen abundance in the
\texttt{vvAPEC} model and varying the helium abundance
implies changing the total density in the model of
a plasma. Therefore, the
change in $\eta$  when fitting this spectrum by the \texttt{APEC}
model reflects mainly the actual change of the density
in the plasma model.

\subsection{The Impact of Thermal Diffusion}
To demonstrate an important role of the thermal
diffusion,
we perform additional calculations by artificially removing
the terms proportional to $r_s$ from Eq.~(\ref{eq:momentum}). In
that case, Eq.~\ref{eq:energy} is not used. These calculations
were performed for NGC~1399 with the highest mean
temperature, $\overline{T} = 0.93$ keV, and NGC~3923 with the
lowest mean temperature, $\overline{T} = 0.56$ keV.

\begin{figure}
\centering
\includegraphics[width=1.05 \columnwidth]{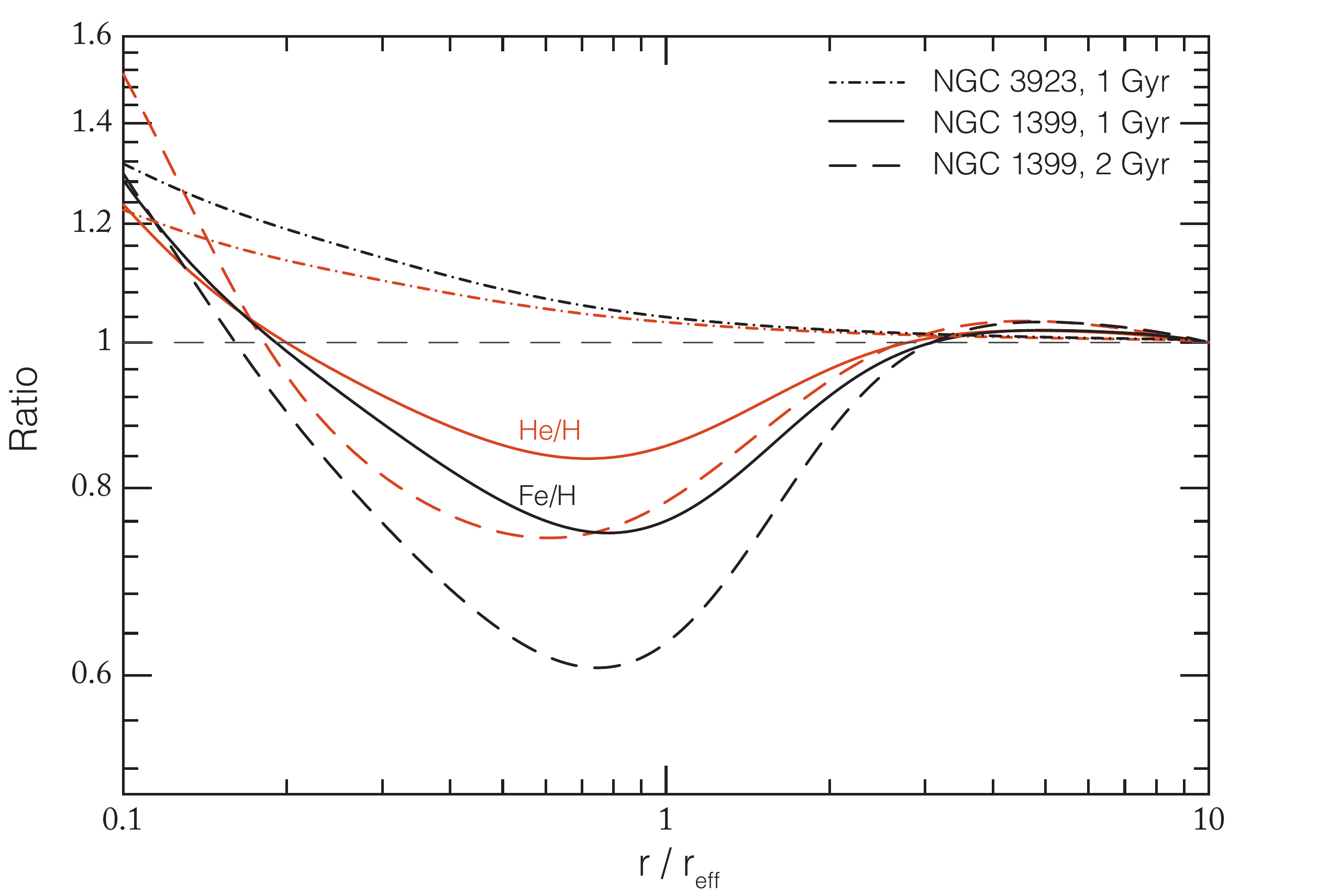}
\caption{Spatial distribution of the \ion{He}{III} (red lines) and \ion{Fe}{XXII} (black lines) abundances calculated with thermal diffusion
divided by the corresponding distribution found without thermal diffusion. The results are shown for 1 (solid lines) and 2 (long dashed lines) Gyr for NGC~1399 and 1~Gyr (dash-dotted lines) for NGC~3923.}
\label{fig:thermaldif}
\end{figure}

Figure~\ref{fig:thermaldif} shows the ratio of the helium (red lines)
and iron (black lines) abundances derived with and
without thermal diffusion (the element abundance is
defined as $n_s/n_{{\rm H}}$). We show the result for NGC~1399
for 1~and 2~Gyr and for NGC~3923 for 1~Gyr. We
see that thermal diffusion changes significantly the
element abundance profile and the amplitude of the
effect. Thermal diffusion exerts the greatest influence
on ions with a large nuclear charge. In addition,
thermal diffusion slows down the sedimentation
of elements heavier than hydrogen for NGC~1399,
whose temperature profile has a positive gradient.
For NGC~3923 the reverse is true: thermal diffusion
accelerates the sedimentation of elements, because
the temperature profile has a negative gradient. The
same is also characteristic of all low-mass galaxies,
where the temperature gradient is negative.

\section{Conclusions}
We calculated the diffusion of elements in the interstellar
medium (ISM) for \Ngal\  early-type galaxies. To estimate
the maximum effect of diffusion, we considered
the full set of Burgers' equations in the approximation
of a non-magnetized ISM plasma using observational
data for the gas density and temperature. Our
calculations showed that a significant gravitational
sedimentation of helium and other heavy elements
is possible. For X-ray-bright galaxies with a rising
radial temperature profile the average relative increase
of the helium mass is 23\% within $1\,\re$ in 1~Gyr. For
less massive galaxies with a flat or declining radial temperature profiles
the corresponding increase in helium is 60\%. The
effect of thermal diffusion accelerates significantly the
sedimentation of elements for galaxies of this type,
while for cool-core galaxies thermal diffusion leads to
a reduction in sedimentation.

We compared our results with the gravitational
helium sedimentation amplitude in cool-core galaxy
clusters. It turned out that the helium abundance in
the ISM could change equally significantly
as in the ICM plasma. This occurs despite the
fact that the temperature of the ISM is
considerably lower than the ICM temperature.
The strong effect of diffusion in the ISM is associated with a lower gas mass fraction
and a complex mass--temperatures relation for early-type
galaxies.

Helium sedimentation may be partially responsible for the
reduced heavy element abundances determined from
X-ray spectroscopy. A two-fold increase in the helium
abundance relative to the solar value leads to an
underestimation of the metal abundance by 20\%.

An interstellar medium with an enhanced helium
abundance can affect the evolution of the galactic
stellar population. Whereas most early-type galaxies
contain mainly an old stellar population, observational
data demonstrate that the star formation process
still takes place in some of such objects \citep[see, e.g.,][]{ODea2008}. 
In particular, in large elliptical
galaxies at the center of cool-core clusters
the star formation rate can reach dozens of$\msun\,\mbox{year}^{-1}$
\citep{ODea2008}. Thus, as a result of sedimentation,
the cooling ISM gas can form helium enriched
stars that leave the main sequence more
rapidly. Therefore, hot stars can fall on the horizontal
branch of the Hertzsprung-Russell diagram more
effectively. Such stars are brighter in the ultraviolet,
which can be responsible for the ultraviolet upturn in
early-type galaxies \citep{Dorman1995,Peng2009a}.

The \Chandra\  and \XMM\  X-ray observatories
have already observed dozens of early-type
galaxies. This allows the physical processes in the
interstellar gas of galaxies to be studied in detail.
Adding the diffusion of elements to the simulations of
such objects can help in forming a clear understanding
of their observational properties.

\section*{Acknowledgements}
P.M. and S.S. thank the Russian Science Foundation
for the support of this work (project no. 14-12-01315).

\bibliographystyle{mnras}
\bibliography{diff_ISM}

%%%%%%%%%%%%%%%%%%%%%%%%%%%%%%%%%%%%%%%%%%%%%%%%%%

%%%%%%%%%%%%%%%%% APPENDICES %%%%%%%%%%%%%%%%%%%%%

%\appendix
%\section*{Profiles}
\clearpage

\begin{figure*}
\centering
\includegraphics[width=0.81\linewidth]{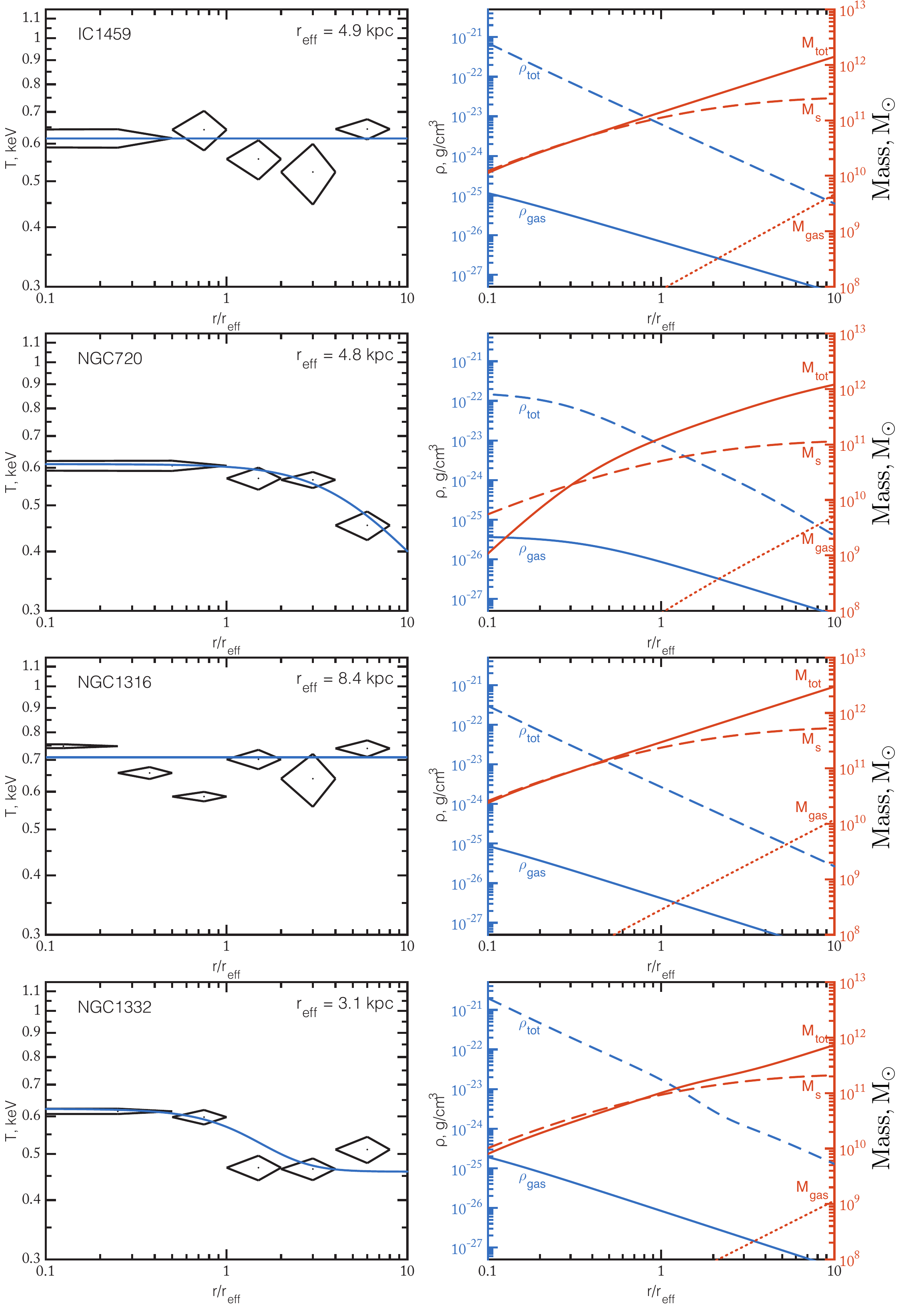}
\caption{
Left: Interstellar gas temperature versus radius (in units of the effective radius) for the galaxies from Table~\ref{tab:sample}. The
diamonds denote the observed temperature with errors; the solid lines indicate the fit from Eqs.~(\ref{eq:T1})--(\ref{eq:T3}). Right: The left vertical
axes correspond to the gas densities ($\rho_{\rm gas}$, the solid line) and the total mass ($\rho_{\rm tot}$, the dashed line) found from Eqs.~(\ref{eq:L})~and~(\ref{eq:M}).
The right vertical axes correspond to the profiles of the total mass ($M_{\rm tot}$ , the solid line), the gas mass ($M_{\rm gas}$, the dotted line),
and the best fit to the stellar component ($M_{\rm s}$, the dash line).}
\label{fig:prof1}
\end{figure*}
\begin{figure*}
\centering
\hspace{1.3cm}
\includegraphics[width=0.85\linewidth]{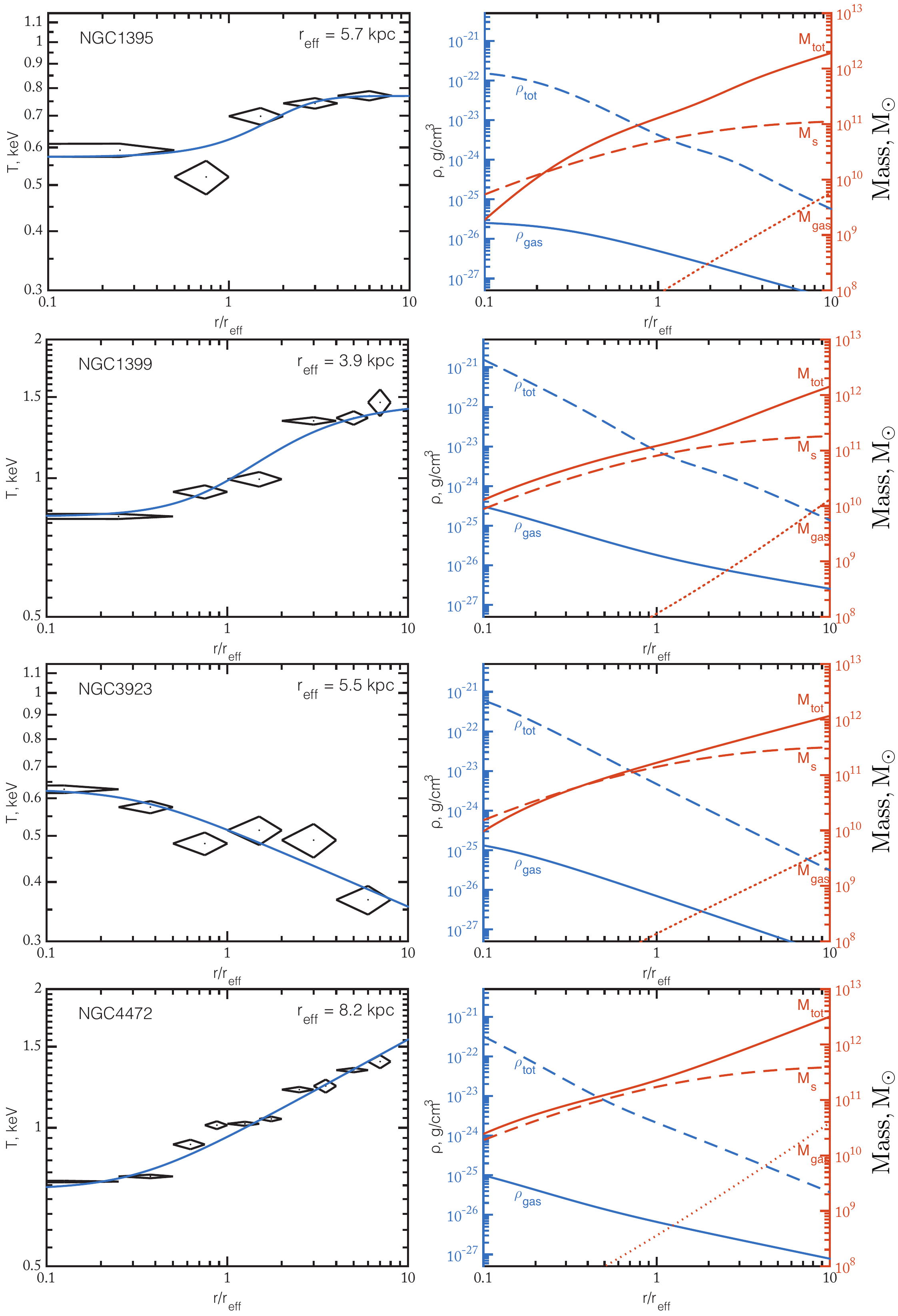}
\contcaption{}
\label{fig:prof2}
\end{figure*}

\begin{figure*}
\centering
\hspace{2cm}
\includegraphics[width=0.75\linewidth]{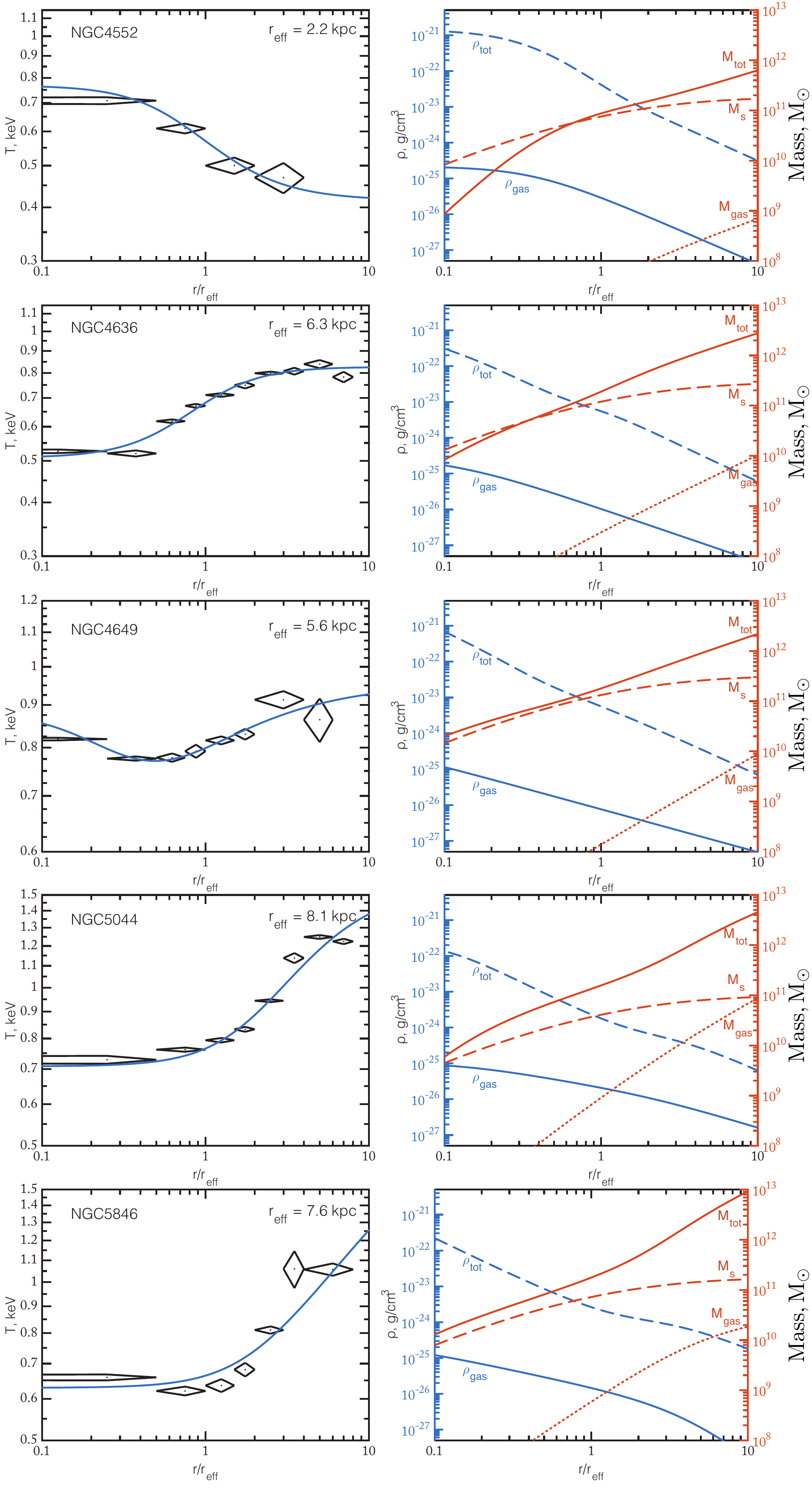}
\contcaption{}
\label{fig:prof3}
\end{figure*}

\clearpage
%\thispagestyle{empty}
%\onecolumn
%\section*{Profiles}
%\begin{landscape}

\begin{table*}
\begin{center}
\caption{Change in the mass fraction of elements in the interstellar gas within $1\,\re$, $4\,\re$ and $10\,\re$ after 1 (2)~Gyr of
diffusion.The change in the mass fraction of an element ($s = $ \ion{He}{III},
\ion{O}{VIII}, \ion{Si}{XIII}, \ion{Fe}{XXII}), divided by the hydrogen mass:
$(\frac{M_{s}}{M_{{\rm H}}} / \left(\frac{M_{s}}{M_{{\rm  H}}}\right)_{t=0} - 1) \times 100$, where the mass is calculated within $1\,\re$, $4\,\re$, and $10\,\re$. The values for 2 and 1~Gyr of diffusion
are given in and without parentheses, respectively.}
\label{tab:results}
\begin{tabular}{lcccc|cccc|c}
\hline
\hline
Galaxy & \ion{He}{III}/H& \ion{O}{VIII}/H & \ion{Si}{XIII}/H & \ion{Fe}{XXII}/H &  \ion{He}{III}/H & \ion{O}{VIII}/H  & \ion{Si}{XIII}/H  & \ion{Fe}{XXII}/H & \ion{He}{III}/H\\
         &    \multicolumn{4}{c}{ $< 1\, \re$} & \multicolumn{4}{c}{$< 4\, \re$} &   $< 10\, \re$    \\
\hline
IC1459 & 51.4 (110.5) & 35.1 (75.3) & 31.4 (67.5) & 29.5 (63.5) & 17.6 (36.4) & 12.0 (24.8) & 10.8 (22.2) & 10.1 (20.9) & 8.1 (15.9) \\
NGC720 & 46.8 (101.9) & 32.8 (71.7) & 29.7 (64.9) & 28.0 (61.3) & 14.8 (30.1) & 11.6 (23.7) & 10.8 (22.1) & 10.4 (21.2) & 4.7 (9.4) \\
NGC1316 & 40.6 (85.8) & 27.7 (58.4) & 24.8 (52.4) & 23.3 (49.2) & 16.0 (33.0) & 10.9 (22.5) & 9.8 (20.2) & 9.2 (18.9) & 8.1 (15.8) \\
NGC1332 & 114.5 (243.2) & 88.5 (188.5) & 82.1 (175.3) & 78.6 (168.1) & 23.4 (48.7) & 16.4 (33.9) & 14.8 (30.6) & 14.0 (28.9) & 11.1 (21.7) \\
NGC1395 & 38.8 (84.0) & 20.5 (43.3) & 16.7 (35.6) & 14.8 (31.7) & 25.8 (54.7) & 17.0 (36.3) & 15.0 (32.1) & 13.9 (29.9) & 12.0 (23.4) \\
NGC1399 & 41.8 (90.0) & 13.4 (28.1) & 7.8 (17.3) & 5.1 (12.1) & 18.8 (39.6) & 8.7 (18.6) & 6.4 (14.0) & 5.3 (11.7) & 8.6 (16.5) \\
NGC3923 & 32.6 (67.2) & 24.5 (50.6) & 22.6 (46.7) & 21.6 (44.7) & 8.6 (17.3) & 6.5 (13.1) & 6.0 (12.1) & 5.8 (11.6) & 3.5 (7.0) \\
NGC4472 & 23.9 (50.2) & 8.6 (18.1) & 5.4 (11.7) & 3.8 (8.6) & 8.7 (17.9) & 1.9 (4.0) & 0.4 (1.1) & -0.3 (-0.4) & 4.3 (8.2) \\
NGC4552 & 88.0 (181.3) & 68.7 (139.7) & 64.0 (130.1) & 61.6 (125.0) & 22.7 (47.3) & 16.2 (33.4) & 14.7 (30.4) & 13.9 (28.8) & 14.1 (27.6) \\
NGC4636 & 17.5 (38.4) & 8.1 (18.5) & 6.2 (14.6) & 5.2 (12.6) & 18.3 (38.3) & 12.1 (25.4) & 10.8 (22.6) & 10.0 (21.1) & 10.7 (21.0) \\
NGC4649 & 55.7 (123.1) & 33.7 (74.2) & 29.0 (64.3) & 26.6 (59.2) & 23.7 (50.0) & 15.1 (32.0) & 13.2 (28.1) & 12.2 (26.0) & 11.3 (21.9) \\
NGC5044 & 5.2 (10.4) & 2.4 (4.7) & 1.7 (3.5) & 1.4 (2.8) & 2.3 (4.8) & 0.1 (0.3) & -0.3 (-0.6) & -0.6 (-1.1) & 3.5 (6.7) \\
NGC5846 & 8.1 (15.3) & 4.7 (8.8) & 4.0 (7.4) & 3.6 (6.6) & 10.7 (26.4) & 5.0 (12.8) & 3.9 (10.4) & 3.3 (9.2) & 25.9 (49.0) \\
\hline

\end{tabular}
\end{center}
\end{table*}
%\end{landscape}

\label{lastpage}
\end{document}